\documentclass[twocolumn, showpacs,preprintnumbers,nofootinbib,prd,superscriptaddress,10pt]{revtex4-1}
\usepackage{graphicx}
\usepackage{xcolor}
\usepackage{lipsum}

\begin{document}
\title{Dynamics of quantum anisotropies in a Taub Universe in the WKB approximation}
\author{Mariaveronica De Angelis}
\email{deangelis.1644941@studenti.uniroma1.it}
\affiliation{Physics Department (VEF), Sapienza University of Rome, P.le A. Moro 5 (00185) Roma, Italy}

\author{Giovanni Montani}
\email{giovanni.montani@enea.it}
\affiliation{Physics Department (VEF), Sapienza University of Rome, P.le A. Moro 5 (00185) Roma, Italy}
\affiliation{ENEA, Fusion and Nuclear Safety Department, C.R. Frascati, Via E. Fermi 45 (00044) Frascati (RM), Italy}
	
\begin{abstract}
We analyze the dynamics of a Taub cosmological model in the presence of a massless minimally coupled scalar field and a cosmological constant, in the limit when both the Universe volume and the scalar field live in a quasi-classical approximation. 
By other words, we study the dynamics of a quantum small anisotropy evolving on a de Sitter background and in the presence of a kinetic term of the inflaton field.

We demonstrate that the quantum anisotropy exponentially 
decays during the Universe expansion, approaching a finite and small value.
This result suggests that the quantum isotropization of 
the Universe during a de Sitter phase is much weaker than the corresponding classical evolution, favouring the survival of certain degree of anisotropy to the de Sitter phase. 
Finally we analyze the case when also the scalar field is considered as quantum variable, by showing how its variance naturally spreads because of no potential term significantly affects its dynamics. This behaviour results to be different from the anisotropy which is subjected to the potential coming out from the spatial curvature. 
\end{abstract}
	
\maketitle
\section{Introduction}
One of the most interesting open questions in theoretical cosmology concerns how a primordial quantum Universe (whose cosmological singularity is intended regularized by a cut-off effect into a Big-Bounce \cite{Ash}) reaches a classical isotropic limit \cite{Primordial}. The reason to hypothesize, near the singularity, a very general morphology of the Universe, relies on the request to address the quantum cosmological problem on a general ground, 
at least within the framework of the Bianchi homogeneous models \cite{Canonical} (we recall that the Bianchi types VIII and IX are prototype for the generic inhomogeneous cosmological problem) \cite{Bel82, Kirillov93, Mont95}. In fact, the isotropic Robertson-Walker model is highly symmetric and it does not contain real gravitational degrees of freedom (actually in cosmology, the two gravitational  degrees of freedom are identified in the anisotropies of space). 
Furthermore, the implementation of symmetry restrictions and the canonical quantization procedure do not commute in
general. 

In \cite{Kir-Mont97} it was argued that, starting with a generic quantum inhomogeneous Universe, it can reaches a classical limit only after it has also became essentially isotropic, otherwise no stable averaged background can emerge. 

While the question concerning how the quantum anisotropies 
can be reduced to small effects is still fully open, in \cite{Battisti} it was shown how such small anisotropies can be naturally damped on a quantum level. This conclusion was mainly based on the features of the basic modes of the associated quantum dynamics. 

Here, we focus our attention to the Taub cosmological model \cite{Taub} in order to deepen and complete the previous study, by analyzing in detail the evolution of wave packets. 

More specifically, we consider a Taub cosmology, in the 
presence of a cosmological constant and a free minimally coupled scalar field (these two last ingredients well mimic the slow-rolling phase of an inflationary scenario \cite{Primordial,Early}). 
The analysis of the dynamics is performed in the semi-classical picture developed in \cite{Vilenkin} to interpret the anisotropy wave function. By other words, we consider the Universe volume and the scalar field as quasi-classical variables while the anisotropy variable contained in the model is fully quantized. 
The potential term appearing in the Taub Hamiltonian is then expanded for small value of the anisotropy, according to the idea proposed in \cite{Vilenkin} that the quantum subsystem must be "small", for a better characterization of this hypothesis see also \cite{Mont-Agost}. 
Validity of the small anisotropy approximation across the
system dynamics is then ensured by analyzing the time dependence of the surviving harmonic potential and of the decaying of a tunnelling process probability toward large values of the anisotropy variable. 

The resulting system is a quantum harmonic oscillator in 
the anisotropy variable, having a frequency rapidly increasing with time. 
The behaviour of Gaussian packets is investigated both via an expansion of the initial condition in terms of the basic modes of the time-dependent harmonic oscillator, as 
well as by an exact Gaussian solution (taken in the 
spirit of \cite{Kiefer2016}). 
Both these studies unambiguously demonstrate that the 
variance of the anisotropy rapidly decreases, reducing to a small but finite non-zero value.
 
This result suggests that if the slow-rolling phase starts when the anisotropy of the Universe is still a quantum degree of freedom, under suitable conditions, it would have chances to survive after the de Sitter phase. Actually, the present analysis also amends for the conclusion in the analysis \cite{Battisti}. In fact, by a refined analytical treatment we demonstrated that the asymptotic limit of the anisotropy standard deviation, for large Universe volume, peaks a small but not zero value.
We also analyzed the classical behaviour of the anisotropy for a comparison for the quantum analyses, also in this case it decays to a constant value but differently from the quantum case this value can be gauged out by a redefinition of the spatial coordinate.

Finally, we consider separately the case when the 
free massless scalar field is quantum too, in order to 
outline that its behaviour is intrinsically different to 
that one of the anisotropy. In fact, the scalar field 
is essentially potential-free during the slow-rolling phase and we see that its quantum variance spreads, suggesting that it is not suppressed by the exponential expansion of the Universe, but it remains a pure quantum degree of freedom. 
It is just such a behaviour that allows the scalar field inhomogeneities (not addressed here) to be the natural origin of the actual Universe clumpiness, while the scalar curvature acts on the anisotropy degrees of freedom so that their evolution is strongly damped.

The paper is structured as follows. In Sec.\ref{II} we give a detailed description of the Taub cosmological model, showing the metric morphology and the associated dynamics. In Sec.\ref{III} we discuss the WKB approach to a small quantum subsystem, expressing a necessary condition for a possible division of the phase-space into a classical and quantum one. In Sec.\ref{IV} we derive the basic equations and solutions to describe the anisotropy dynamical evolution, showing its behaviour during a de Sitter phase. Sec.\ref{V} is devoted to analyze the behaviour of probability density of the anisotropy variable as the Universe expands from the singularity, building a complete wave function and studying the wave packets with a Gaussian Ansatz. In Sec.\ref{VI} we show the behaviour of the anisotropy considered as a classical degree of freedom. In Sec.\ref{VII} we justify the small oscillation assumption for the considered model. In Sec.\ref{VIII} we investigate quantum scalar field fluctuations showing how they can survive to the de Sitter phase producing seeds. Finally in Sec.\ref{IX} conclusions are drawn.

\section{The Taub model}\label{II}

The Taub cosmological model is an homogeneous Universe. 
The presence of a different evolution of a scale factor from the other two makes this model anisotropic. For this reason, Taub Universe is invariant around rotation about one axis of three-dimensional space.

The line element of the space-time reads as
\begin{equation}
ds^2=N^2(t)dt^2-e^{2\alpha}(e^{2\beta})_{ab}\omega^a\omega^b,
\end{equation}
where $\omega^a=\omega_{i}^adx^{i}$ are the left-invariant one-forms. 
The variable $\alpha(t)$ describes the isotropic expansion of the model and the gravitational degrees of freedom of the Universe are associated to $\beta_+$, the anisotropy. It is determined in the following traceless symmetric matrix
\begin{equation}
\beta_{ab}=diag(\beta_+,\beta_+,-2\beta_+).
\end{equation}
In addition, Taub model is a particular case of Bianchi IX model once $\beta_- \equiv0$.
We introduce the cosmological constant $\Lambda$ because we want to describe the de Sitter phase and a scalar field $\phi$. 
The behaviour of a massless scalar field well approximates that one of an inflation field during the slow-rolling dynamics, $\dot{\phi^2} \ll |V(\phi)|$, when the potential term is essentially constant and it provides the cosmological term.

We adopt natural units $\hbar=c=1$ apart from where the classical limit is discussed.

The dynamics of the Taub model is described by the action $I$ with the Misner variables

\begin{equation}
I=\int dt(p_{\alpha}\partial_th^{\alpha}-N\mathcal{H}),
\label{action}
\end{equation}
in which $N(t)>0$ is the lapse function, $h^{\alpha}$ is an unified notation for minisuperspace variables (i.e. $h_{ab}$ and $\phi$, $\alpha=1,2$) and $p_{\alpha}$ the conjugate momenta of $h^{\alpha}$. The variation with respect to $N$ generates the scalar constraint $\mathcal{H}=0$ which reads as
\begin{equation}
\mathcal{H}=\frac{k}{3(8\pi)^2}e^{-3\alpha}(-p_{\alpha}^2+p_+^2+p_{\phi}^2+\mathcal{V}+\Lambda e^{6\alpha})=0,
\end{equation}
in which $k=8\pi G$ is the Einstein constant and the potential $\mathcal{V}$ takes the form
\begin{equation}
\mathcal{V}\equiv-\frac{6(4\pi)^4}{k^2}\eta \ ^3R=\frac{3(4\pi)^4}{k^2}e^{4\alpha}U(\beta_+)
\end{equation}
where the spatial scalar of curvature generates Taub potential term $U(\beta_+)$
\begin{equation}
U(\beta_+)=e^{-8\beta_+}-4e^{-2\beta_+}.
\end{equation}
The conjugate momenta expression can be constructed by the inversion of the relation coming from the first of the Hamilton equations
\begin{equation}
p_{\alpha}= -\frac{6(4\pi)^2}{Nk}e^{3\alpha}\dot{\alpha}
\label{p}.
\end{equation}

Adopting the change of variable $a(t)$ as $e^{\alpha}$, $\mathcal{H}=0$ becomes
\begin{equation}
\mathcal{H}= k\biggl[-\frac{p_a^2}{a}+\frac{p_+^2+p_{\phi}^2}{a^3} \biggl] + \frac{a U(\beta_+)}{4k}+\frac{\Lambda a^3}{k}=0,
\end{equation}
in which $p_{\phi}$ is a constant of motion because of the absence of a potential term $V(\phi)$. The phase space of this system is six-dimensional with coordinates $a,p_a,\beta_+, p_+,\phi, p_{\phi}$. The dynamical picture is completed by taking into account the choice $N=a^3/k$ which fixes the temporal gauge. The cosmological singularity appears as $a \rightarrow 0$. Far from the singularity, the cosmological constant term dominates over the scalar fields kinetic energy and it is necessary for the development of the inflationary scenario.

\section{Vilenkin approach to the small quantum subsystem} \label{III}

In quantum cosmology, the wave function of the Universe is a functional defined on the minisuperspace metric $h_{ab}(x)$, i.e.  
\begin{equation}
\psi (h_{ab}(x)).
\end{equation}
We stress that an external time definition is absent because of the null scalar constraint $\mathcal{H}=0$. 
In this perspective, we can consider a small quantum subsystem of the semi-classical Universe. Hence, the Hamiltonian reads as
\begin{equation}
\mathcal{H}=H_0+H_q.
\end{equation}
We also assume that the quantum variables $q^{\nu}$ with $\nu=(1, ..,\ n-m)$ does not effect the dynamics of the classical ones $h^{\alpha}$ with $\alpha=(1, ..,\ m)$ which is a Wentzel-Kramers-Brillouin approach and the Born-Oppenheimer approximation.

The Wheeler-DeWitt equation corresponding from the action (\ref{action}) can be written as follows
\begin{equation}
(\nabla_0^2-U_0-H_q)\psi=0,
\label{wdw}
\end{equation}
in which the operator $H_0=\nabla_0^2-U_0(h)$ represents the classical Hamiltonian obtained by neglecting the quantum variables and the respective momenta $p_{\nu}=-i\partial/\partial q^{\nu}$. To justify the smallness of the quantum subspace, Vilenkin \cite{Vilenkin} imposed the following reasonable assumption
\begin{equation}
\frac{H_q\ \psi}{H_0\ \psi}= O(\hbar),
\end{equation}
so that 
\begin{equation}
\nabla_q^2=O(\hbar^{-1}).
\end{equation}
The necessary condition for a possible division of the full phase-space into a classical and a quantum subsystem is that the minisuperspace metric tensor $g_{ab}^{(0)}\sim 1$ and $g_{ab}=O(\hbar)$.

The wave function of the Universe can be written as
\begin{equation}
\psi(h,q)= \psi(h)\chi(h,q),
\end{equation}
where $\psi(h)=A(h)\ e^{iI(h)}$.

In such a way, equation (\ref{wdw}) can be decomposed in three equations in order of $\hbar$. In the lowest order we obtain the Hamilton-Jacobi equation for the classical action $I$ and, in the next order, an equation for the amplitude $A$ which takes the form of a continuity equation. They respectively are
\begin{equation}
g^{ab}(\nabla_aI)(\nabla_bI)+U=0
\label{HJ}
\end{equation}
\begin{equation}
2\nabla A \cdot \nabla I+A\nabla^2I=0.
\label{cont}
\end{equation}

The equation for the wave function $\chi(h,q)$ of the quantum subspace at the same order in $\hbar$ of (\ref{cont}) has the form
\begin{equation}
2i(\nabla_0I)\nabla_0 \chi=H_q\chi,
\label{s}
\end{equation}
derived by decoupling it via the adiabatic approximation expressed by the condition $|\partial_hA(h)|\gg|\partial_h\chi(h,q)|$.
Using the Hamilton-Jacobi equation, (\ref{s}) can be rewritten as
\begin{equation}
i\frac{\partial \chi}{\partial \tau}=H_q\chi,
\end{equation}
with $d\tau=N(t)dt.$ Hence, the Schr\"odinger equation we find for the subsystem in the background defined by $h^{\alpha}$ allows to define a dynamical evolution for the quantum subspace. In both follows, the minisuperspace variables division between $h^{\alpha}$ and $q^{\nu}$ corresponds to the following: the volume of the Universe and the scalar field are taken as classical variables, while the anisotropy $\beta_+$ is regarded as the quantum one.

The total probability density is defined by the wave function $\psi=A(h)e^{iI(h)}\chi(h,q)$ and corresponds to a conserved current. It is the product of the classical and the quantum part
\begin{equation}
\rho(h,q,t)=\rho_0(h,t)|\chi(q,h(t),t)|^2
\end{equation}
in which $\rho_0(h,t)=|A(h)|^2$ and it is normalized by
\begin{equation}
\int \rho_0 d\Sigma_0=1,
\end{equation}
where $d\Sigma_0$ is the surface element in the subsystem defined by $h^{\alpha}$ and $\chi(q,h,t)$ can be normalized by
\begin{equation}
\int |\chi|^2 d\Omega_q=1
\end{equation}
in which $d\Omega_q=|detg_{\mu \nu}|^{1/2}d^nq.$ This is the standard interpretation of the wave function for a small subspace of the Universe.

\section{Basic equations and solutions for the Taub model}\label{IV}
To describe the dynamical evolution of the Taub model, we need to analyze the three equations derived by Vilenkin approach.
Equation (\ref{HJ}) and (\ref{cont}) become
\begin{equation}
-(\partial_{\alpha}I)^2+(\partial_{\phi}I)^2 +\Lambda e^{6\alpha}=0,
\label{I}
\end{equation}
\begin{equation}
\partial_{\alpha}(A^2\partial_{\alpha}I)+\partial_{\phi}(A^2\partial_{\phi}I)=0.
\label{A}
\end{equation}
It has been used the notation $e^{\alpha}$ to simplify the analytical integration.\\
From (\ref{I}) we construct the classical dynamics, corresponding to the zero order in $\hbar$. In solving (\ref{I}), we see that the implementation of the
standard Hamilton-Jacobi method suggests that $\partial_{\phi}I=p_{\phi}=const$ (this fact reflects the massless free nature of the scalar field
at the zero order in $\hbar$) and the solution can be expressed with the ansatz
\begin{equation}
I(\alpha, \phi)=f(\phi)+g(\alpha),
\end{equation}
and reads as
\begin{equation}
I(\alpha, \phi)=p_{\phi}\phi+\tilde{I}(\alpha),
\end{equation}
where
\begin{equation}
\tilde{I}(\alpha)=\mp\frac{1}{3}\sqrt{\Lambda e^{6\alpha}+p_{\phi}^2}\pm\frac{p_{\phi}}{3}\ arctanh\biggl[\frac{\sqrt{\Lambda e^{6\alpha}+p_{\phi}^2}}{p_{\phi}}\biggl]+\ c,
\end{equation}
with $c$ integration constant.

From the continuity equation (\ref{A}) it has been found the amplitude $A$ by variables separation
\begin{equation}
A(\alpha, \phi))=A_1(\alpha)A_2(\phi).
\end{equation}
As a result, a simple solution of this non-linear equation is
\begin{equation}
A(\alpha, \phi)=\frac{e^{-\frac{c}{6p_{\phi}}arctanh\biggl(\frac{\sqrt{\Lambda e^{6\alpha}+p_{\phi}^2}}{p_{\phi}}\biggl)-\frac{c}{2p_{\phi}}\phi}}{(\Lambda e^{6\alpha}+p_{\phi}^2)^{1/4}}
\end{equation}
with $c$ parameter of the variables separation.\\

The functions $I$ and $A$ provide a complete characterization of the quasi-classical system.

\subsection{Time-dependent harmonic oscillator}
To describe the behaviour of the anisotropy $\beta_+$ we now study equation (\ref{s}), that is a pure Schr\"odinger-like equation 
\begin{equation}
2i\ \frac{e^{3\alpha}}{Nk}\biggl(\dot{\alpha}\frac{\partial \chi}{\partial \alpha}+\dot{\phi}\frac{\partial \chi}{\partial \phi}\biggl)=H_q \chi,
\label{schro}
\end{equation}
where $\chi=\chi(\alpha(t),\phi(t),\beta_+)$.
Hence, using the previously introduced change of variables $e^{\alpha}=a$ and the fixed temporal gauge, (\ref{schro}) can be rewritten as
\begin{equation}
i\partial_{\tau} \chi=H_q \chi,
\label{schro2}
\end{equation}
in which the quantum Hamiltonian $H_q$ reads as
\begin{equation}
H_q=-\frac{\partial^2}{\partial\beta_+^2}+\frac{a^4}{4k^2}U(\beta_+).
\label{Hq}
\end{equation}
We highlight that the variable $\alpha$ increases with the synchronous time while $\tau$ decreases. In this respect, we have
\begin{equation}
\frac{d\alpha}{d\tau}= -2kp_{\alpha}<0,
\label{alpha}
\end{equation}
with $p_{\alpha}\sim\sqrt{\Lambda e^{6\alpha}}$.\footnote{In (\ref{I}), $p_{\phi}^2$ can be neglected for high values of $\alpha$. We stress that, since we are considering an expanding Universe in the adopted time variable we must take the positive square root when solving the $\alpha$ dependence.} If we solve (\ref{alpha}), we get $\tau=a^{-3}/(6k\sqrt{\Lambda})$. To show the behaviour of $\tau$ compared to the new variable $a$, we compute
\begin{equation}
\frac{d\tau}{dt}=-\frac{1}{2k\sqrt{\Lambda}a^4}\frac{da}{dt}.
\end{equation}

Moreover, according to the Vilenkin idea of a small quantum system (see also \cite{Mont-Agost}), we consider the quasi-isotropic regime $|\beta_+| \ll 1$ so that the potential term gets a quadratic form
\begin{equation}
U(\beta_+)=-3+24\beta_+^2,
\end{equation}
in which the zero order of the approximate potential, substituted in WDW (i.e. $-3e^{4\alpha}\equiv -3a^4$) would provide a contribution to the Hamilton-Jacobi equation (\ref{I}) and becomes negligible when the cosmological constant dominates.
Instead for the equation (\ref{schro2}) we get
\begin{equation}
i\partial_{\tau}\chi=\biggl(-\frac{\partial^2}{\partial\beta_+^2}+\omega^2(\tau)\beta_+^2\biggl)\chi,
\label{schroom}
\end{equation}
in which the frequency term is $\omega^2(\tau)=6\tau^{-4/3}/\tilde{k}^2$, and $\tilde{k}^2=k^2 (6k\sqrt{\Lambda})^{4/3}$.

Harmonic oscillator quantum theory with time dependent frequency is known and the solution to (\ref{schro2}) can be obtained analytically by using the exact invariant method and some transformations \cite{Lewis68, Lewis69, Pedrosa}. An exact invariant $J(\tau)$ is a constant
of motion (namely $J^{\prime}\equiv dJ/d\tau = \partial_{\tau}J - i[J, \hat{H_q}]
= 0)$, is hermitian and for the Hamiltonian $H_q$ it explicitly reads

\begin{equation}
J_{+}=\frac{1}{2}(\rho^{-2}\beta_{+}+(\rho p_{+}-\dot{\rho}\beta_{+})^2),
\end{equation}
in which $\rho=\rho(\tau)$ is the function satisfying the auxiliary differential equation
\begin{equation}
\ddot{\rho}+\omega^2\rho-\rho^{-3}=0.
\label{rho}
\end{equation}

The solution is connected to the J-eigenfunctions $\psi_n$ by the relation $\chi_n(\beta_+,\tau)=e^{i\alpha_n(\tau)}\psi_n(\beta_+,\tau)$ but the general one is a linear combination $\chi(\beta_+, \tau)=\sum_nc_n\chi_n(\beta_+,\tau)$, in which $c_n$ are real or complex coefficients that weight the different wave functions. $\chi_n$ reads as
\begin{equation}
\chi_n(\beta_+,\tau)=\frac{ e^{i\alpha_n(\tau)}}{\sqrt{\sqrt{\pi}\ n!\ 2^n \rho}}\; h_n\biggl(\frac{\beta_+}{\rho}\biggl)\; e^{[\frac{i}{2}(\frac{\dot{\rho}}{\rho}+i\frac{1}{\rho^2})\beta_+^2]}.
\label{eigen}
\end{equation}
In (\ref{eigen}) $h_n$ are Hermite polynomials and the phase $\alpha(\tau)$ is given by
\begin{equation}
\alpha_n=-\biggl(n+\frac{1}{2}\biggl)\int \frac{d\tau}{\rho^2(\tau)}.
\end{equation}

The non-trivial step in this construction is to obtain an analytical solution of the auxiliary equation for $\rho$.\\In this respect, we make use of the method in \cite{Lewis68}. In fact, by writing the most general such invariant in terms of two independent and linearly solutions (i.e. $h(\tau)$ and $r(\tau)$) of
\begin{equation}
\frac{d^2 q}{d\tau^2}+\omega^2(\tau)q=0,
\end{equation}
which gives the motion in a straight line of a harmonic oscillator, it is possible to write the general solution of the non linear equation
for $\rho$
\begin{equation}
\rho=(\mathcal{W})^{-1}(A^2r^2+B^2h^2+2(A^2B^2-(\mathcal{W})^2)^{1/2}hr)^{1/2},
\end{equation}
where $A^2$, $B^2$ are arbitrary real constants.\\
\\
Hence, we have
\begin{equation}
h(\tau)=\frac{\biggl(\sqrt{6}\cos[\frac{3\sqrt{6}\tau^{1/3}}{\tilde{k}}]+18\tau^{1/3}\sin[\frac{3\sqrt{6}\tau^{1/3}}{\tilde{k}}]\biggl)}{\sqrt{6}\tilde{k}},
\end{equation}

$$
r(\tau)= \frac{\sqrt{\frac{3}{2}\biggl(-18\tau^{1/3}\cos[\frac{3\sqrt{6}\tau^{1/3}}{\tilde{k}}]+\sqrt{6}k \sin[\frac{3\sqrt{6}\tau^{1/3}}{\tilde{k}}]\biggl)}}{8\tilde{k}}
$$
and the Wronskian is
\begin{equation}
\mathcal{W}=h  r^{\prime}-r h^{\prime} = \frac{81 \sqrt{\frac{3}{2}}}{2\tilde{k}^3}.
\end{equation}
By substituting our results of $h(\tau)$ and $r(\tau)$, we obtain
\begin{widetext}
\begin{equation}
\rho(\tau)=\frac{\tilde{k}^3}{324\sqrt{3}}\biggl(\frac{1}{\tilde{k}^2}\biggl( (9A^2+64B^2)(\tilde{k}^2+54\ \tau^{2/3})+\biggl(-9A^2(\tilde{k}^2-54\ \tau^{2/3})+64B^2(\tilde{k}^2-54\ \tau^{2/3})-144\sqrt{24A^2B^2-\frac{59049}{k^6}}\tilde{k}\tau^{1/3}\biggl)
\label{rhoan}
\end{equation}
$$
\cos[\frac{6\sqrt{6}\tau^{1/3}}{\tilde{k}}]+6\sqrt{2}\biggl(2\sqrt{8A^2B^2-\frac{19683}{\tilde{k}^6}}\tilde{k}^2+\sqrt{3}(-9A^2+64B^2)\tilde{k}\tau^{1/3}-108\sqrt{8A^2B^2-\frac{19683}{\tilde{k}^6}}\ \tau^{2/3}\biggl)\sin[\frac{6\sqrt{6}\ \tau^{1/3}}{\tilde{k}}]\biggl)\biggl)^{1/2}.
$$
\end{widetext}

The above scheme allows us to analyze the evolution of the wave function once assigned a generic initial condition.

\section{Analysis of the wave packets}\label{V}
In this section, we analyze the evolution of small quantum subsystem in correspondence to a Gaussian initial condition for the probability distribution of $\beta_+$.\\
The probability density for a generic expansion takes the form
\begin{equation}
|\chi(\beta_+,\tau)|^2\propto|\sum_n c_n \chi_n(\beta_+,\tau)|^2.
\label{chi}
\end{equation}

We consider the following initial condition 
\begin{equation}
\chi_i(\beta_+,\tau_i)=D\ e^{-\beta_+^2/2\sigma_i^2},
\label{psi}
\end{equation}
where $D$ is a normalization constant (i.e. $D$= $1/\sqrt{2 \pi \sigma_i}$). By doing this, we can calculate $c_n(\tau_i)$ and their evolution
\begin{equation}
c_n=\int d\beta_+ \chi_n(\beta_+,\tau)\ \chi_i(\beta_+,\tau_i).
\end{equation}
To build the complete probability distribution we calculate $|\chi_n(\beta_+,\tau_i)|^2$ with $n=(0,\ ..,\ 35)$ terms of Hermite polynomials. Then, in order to show the time evolution, we compute the wave function at different values of $\tau$.

\begin{figure}[h!]
	\centering
	\includegraphics[width=8.7cm]{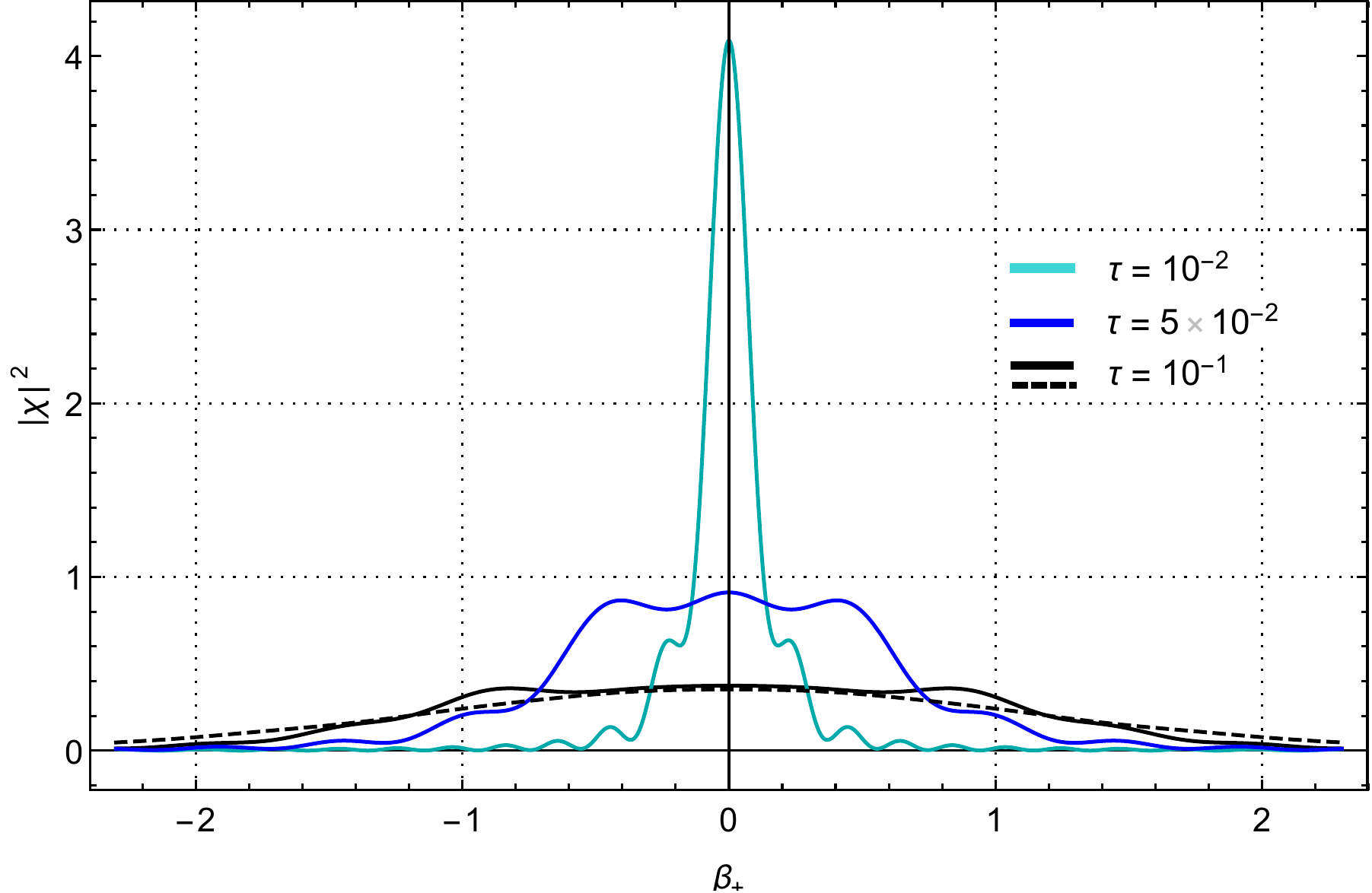}
	\protect\caption{Time evolution of $|\chi(\beta_+,\tau)|^2$ is highlighted by different colours. The considered initial time is $\tau_i$ and we show the difference between the square absolute value of the initial condition (\ref{psi}) (dashed line) and the solution with Hermite polynomials (continuous line). The not well defined behaviour by the wavy trend is given by the truncation of the Hermite polynomials. In the plot we take $A=\frac{81\sqrt{3/2}}{2B}$ and $B=1$.}
	\label{campane}
\end{figure}

We see (Fig. \ref{campane}) that the profile of the Gaussian probability density becomes more
and more peaked as the volume of the Universe expands during the de Sitter phase, i.e. as $\tau \rightarrow 0$.
However, the analytical behaviour we fixed for the function $\rho$ suggests that the Gaussian can not reaches a real delta function as indicated in \cite{Battisti}, while it must emerge a steady small, but finite value of the standard deviation of the quantum variable $\beta_+$
\begin{equation}
\rho(\tau\rightarrow 0)= \frac{2\sqrt{\frac{2}{3}}}{81}\tilde{k}^3+\frac{2}{3}\tilde{k} \tau^{2/3}-\frac{3\sqrt{6}\tau^{4/3}}{\tilde{k}}+O(\tau)^{5/3}.
\label{rhoexp}
\end{equation}
This behaviour is also confirmed by the asymptotic study of an exact Gaussian solution of the time-dependent Schr\"odinger equation, we will discuss below.
In other words, the de Sitter exponential expansion of the Universe strongly depresses the quantum Universe anisotropy, but a small relic standard deviation survives also at the end
of the inflation. This a bit surprising result suggests that, as we shall see below, although the anisotropy can not have the same non-suppressed
behaviour of a scalar field (considered the source of inhomogeneous fluctuations at the ground of the structure formation across the Universe), a small tensor degree of freedom can also remain present on
a quantum level, which in the full inhomogeneous scenario could originate a smaller tensorial component of the primordial spectrum.
Also on a classical level, the Universe anisotropy decays to a constant value, but it has no physical meaning since it can be reabsorbed into the
definition of the 1-forms of the Taub model, differently from the relic
quantum fluctuating one. 

Since it is evident from the harmonic oscillator eigenfunction that the simplest way to locate the Universe is a Gaussian shape, we now search for an exact solution of the time-dependent Schr\"odinger equation \cite{Kiefer2016} as
\begin{equation}
\chi(\beta_+,\tau)=N(\tau)e^{-\frac{1}{2}\Omega(\tau)\beta_+^2}.
\end{equation}
We insert this ansatz in (\ref{schroom}) and separating all the terms with $\beta_+$ of zero and quadratic order, we get the following two equations
\begin{equation}
iN'(\tau)=\frac{1}{2}N(\tau)\Omega(\tau),
\end{equation}
\begin{equation}
i\Omega'(\tau)=(\Omega(\tau))^2-\omega^2(\tau).
\label{gwidth}
\end{equation}
In addition, we also request a normalized wave function for any given time and this provides the modulus of the normalization factor
\begin{equation}
|\chi(\beta_+,\tau)|^2=|N|^2 \int_{-\infty}^{+ \infty}e^{-\frac{1}{2}(\Omega^*+\Omega)\beta_+^2}d\beta_+
\end{equation}
$$
=|N|^2\frac{\sqrt{\pi}}{\sqrt{\Re(\Omega)}}\equiv 1.
$$
It is enough to solve (\ref{gwidth}) for the inverse Gaussian width, to get the physically information on the behaviour of the anisotropy.

If we separate $\Omega$ in its real and complex part respectively, i.e. we
set $\Omega = f(\tau) + i g(\tau)$, then (\ref{gwidth}) provides the following non-linear system
\begin{eqnarray}
2g = \frac{f^{\prime}}{f}
\, , \label{mv1a}\\
g^{\prime} = g^2 + \omega ^2 - f^2.
\label{mv1b}
\end{eqnarray}
These two coupled equations do not admit a simple analytical solution, but
we are interested to the limit $\tau\rightarrow 0$ and we can easily construct the asymptotic behaviour.
In fact, if we set the condition
\begin{equation}
g^{\prime} = \omega ^2 (\tau) =
\frac{c^2}{\tau^{4/3} }
\Rightarrow
g(\tau \rightarrow 0)\simeq - \frac{3c^2}{\tau^{1/3}}
\, ,
\label{mv2}
\end{equation}
where $c^2=6/\tilde{k}^2$, then for the real part of $\Omega$ from (\ref{mv1a}) we get the following asymptotic expression
\begin{equation}
f(\tau\rightarrow 0) \simeq
f_0 e^{- 6 c^2 \tau^{2/3}}
\label{mv3}
\end{equation}
where $f_0$ is an integration constant.

Since the standard deviation $\sigma$ of the
Gaussian probability distribution is
$1/\sqrt{\Re(\Omega)}$, we have
\begin{equation}
\sigma (\tau\rightarrow 0) \simeq
\frac{1}{\sqrt{f_0}}\ e^{\frac{6}{2}c^2\tau^{2/3}}.
\label{mv}
\end{equation}

\begin{figure}[h!]
	\centering
	\includegraphics[width=8.7cm]{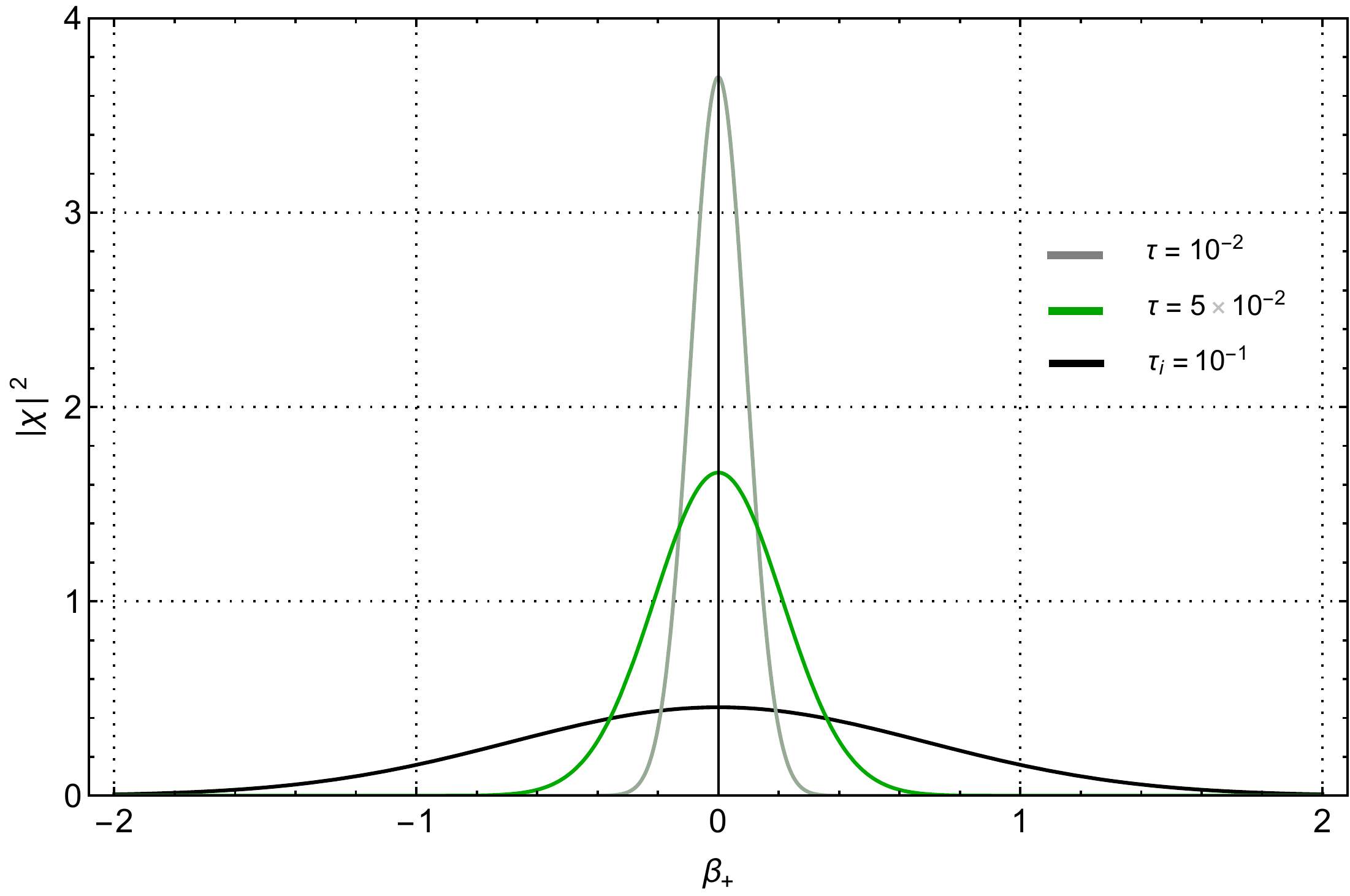}
	\protect\caption{Time evolution of $|\chi(\beta_+,\tau)|^2$ is highlighted by different colours. The considered initial time is $\tau_i$. In the plot we take $\frac{1}{\sqrt{f_0}}=\frac{2\sqrt{2/3}}{81}$.}
	\label{campane2}
\end{figure}

Hence, when the Universe expands
(i.e. $\tau\sim 1/a^3$ decreases) the standard deviation exponentially decays (see Fig. \ref{campane2}). However, the salient point is that it has to approach a constant value.  From the point of view of an exact Gaussian solution, this feature corresponds to the constant value assumed by the function $\rho$ when $\tau$ approaches zero, as in (\ref{rhoexp}).

This study confirms the idea that, although it is in principle a small value, the quantum anisotropy approaches a finite non-zero limit even
after a de Sitter phase.
By other words, if the Universe anisotropy is small enough to be in a quantum regime when inflation starts, it is still present in the late Universe.

The non-vanishing behaviour of the standard deviation of $\beta
_+$, in the limit $\tau \rightarrow 0$ could seem a natural implication of
the Heisenberg uncertainty principle,
since $p_+$ can not diverge (actually, in the considered scheme, it must
remain small, see \cite{Mont-Agost}). However, the frequency of the
considered harmonic oscillator is diverging as $\omega \sim \tau ^{-2/3}$
when $\tau \rightarrow 0$.
Thus, $\omega \langle \beta_+^2\rangle^{1/2}$ can remain small and
positive for $\tau\rightarrow 0$ even in case when the standard deviation
approaches zero ($\omega \langle \beta_+^2\rangle^{1/2}$ is the right variable to be addressed in comparison with a
time-independent harmonic oscillator, having a constant frequency).
The request that the addressed quantum subsystem remains ``small" in the
sense discussed in \cite{Mont-Agost, Vilenkin} leads to link the
limiting small value of the $\beta_+$
standard deviation to the value $\tau _f$ when the de-Sitter phase ends,
i.e. $(\tau _f)^{-2/3}\ll \sqrt{f_0}$.

\section{Classical anisotropy behaviour} \label{VI}

To better understand if only the quantum anisotropy will survive in the late Universe, in this section we will analyze the behaviour of the classical one.

To find the explicit expression for $\beta_+^c(\tau)$, it has been calculated by the Hamilton equation 
\begin{equation}
\frac{\partial\beta_+^c}{\partial \tau}=\frac{\partial \mathcal{H}}{\partial p_+}.
\end{equation}
As a result
\begin{equation}
\beta_+^c(\tau)=2k p_+\tau+\beta_0,
\end{equation}
where the integration constant $\beta_0$ can be set equal to zero by redefinition of the space coordinates.
In this respect, in the limit of the expanding Universe, differently from its quantum behaviour, it is associated to a vanishing value after the de Sitter phase.

\section{The potential as an attractor}\label{VII}
As we see from (\ref{Hq}), the time-dependent frequency term is multiplied in the exact Hamiltonian by the Taub potential. Hence, we get a potential term which changes its shape with the Universe expansion
\begin{equation}
U(\tau, \beta_+)=\frac{1}{\tau^{4/3}}\ U(\beta_+),
\end{equation}
becoming an attractor with a remarkable restoring force. We obtain an increase in depth and width of the potential well as time goes to zero.
We retained the only large contribution for $\beta_+ \rightarrow \infty$, when
\begin{equation}
-\frac{1}{\tau^{4/3}}\ e^{-2\beta_+} \ll 1,
\end{equation} 
to validate the theory of small oscillations approximation of the potential.

\begin{figure} [h!]
	\includegraphics[width=8.5cm]{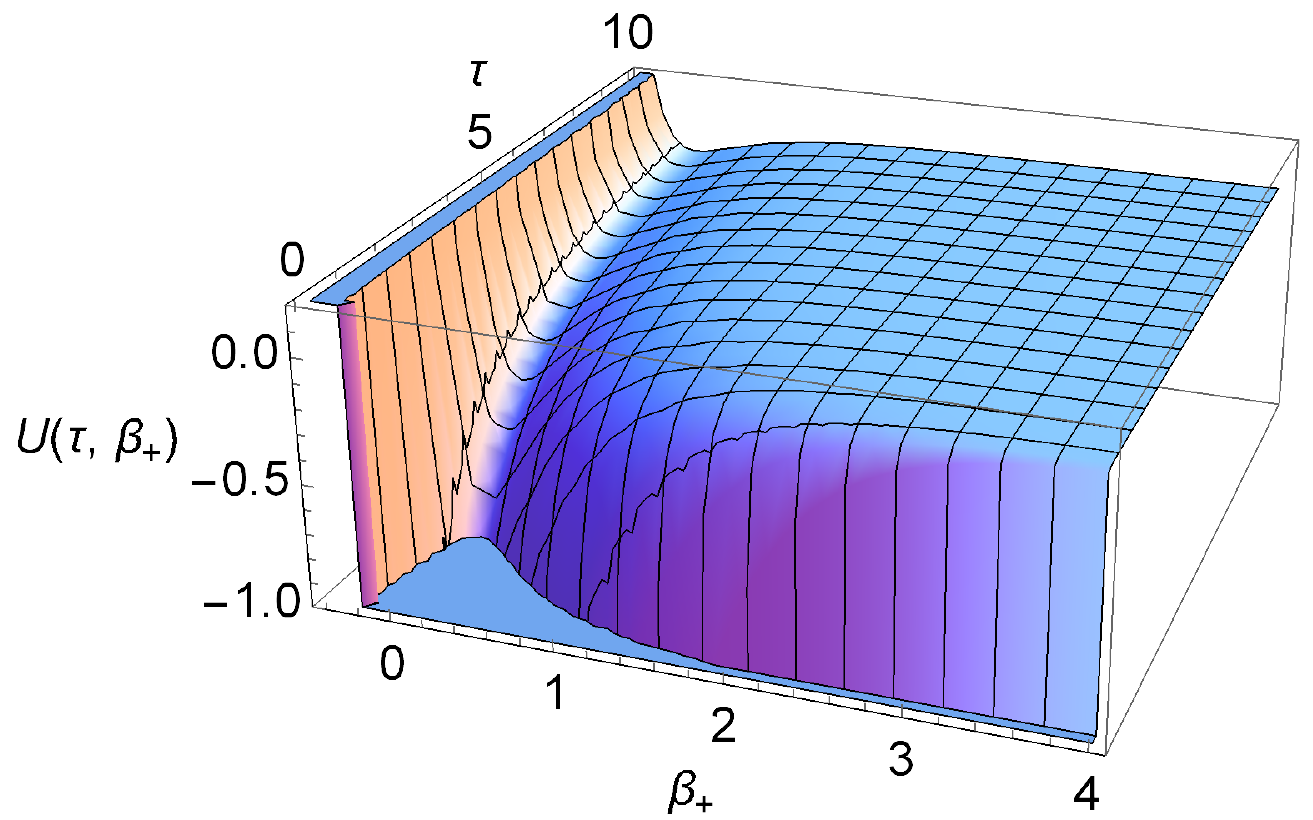}
	\caption{The 3D graph shows the time evolution of the potential $U(\tau, \beta_+)$.}
	\label{pot}
\end{figure}

In term of $\beta_+$ the condition above reads as
\begin{equation}
\beta_+ \gg -\frac{2}{3} \ln \tau.
\end{equation}
Since this condition holds for any values of $\beta_+$ as far as $\tau$ approaches zero, we see that the small oscillation model is a very reliable paradigm for the present analysis (see Fig.\ref{pot}).

\section{Behaviour of the quantum scalar field} \label{VIII}
The cosmological classical field, responsible for the inflation has also small inhomogeneous quantum fluctuations and the generation of density inhomogeneities relies just in considering such small quantum corrections during the slow rolling phase as the sources of the perturbations observed today. In fact, our analysis demonstrated that a quantum anisotropy degree of freedom would be strongly damped during the de Sitter phase.

Since the addressed model is intrinsically homogeneous, we can not consider here the spatial dependence of the quantum field, none the less we can study the case in which also the scalar field is a quantum degree of freedom and compare its behaviour with that one of the anisotropy. Thus, now the classical Hamiltonian contains only the time variable $a$ while the quantum subsystem Hamiltonian reads as
\begin{equation}
H_q=\biggl[-\biggl(\frac{\partial^2}{\partial \phi^2}+\frac{\partial^2}{\partial\beta_+^2}\biggl)+\ U(\beta_+)\biggl]
\end{equation}
leading to the Schr\"odinger equation
\begin{equation}
i\partial_{\tau}\chi=\biggl(-\frac{\partial^2}{\partial\phi^2}+H^{AO}_q\biggl)\chi,
\label{schrophi}
\end{equation}
in which $H^{AO}_q$ refers to (\ref{Hq}). To solve (\ref{schrophi}), we take the following wave function
\begin{equation}
\chi(\beta_+, \tau)=e^{-ip_{\phi}^2\tau}\xi(\beta_+)
\end{equation}
in which a phase factor in $\phi$ is added. It is easy to check that the function $\chi(\beta_+, \tau)$ still satisfies (\ref{schro2}). Moving to the general solution, we get
\begin{equation}
\Psi(\beta_+,\phi,\tau)=\int \frac{dp_{\phi}}{2\pi}e^{-ip_{\phi}^2\tau}e^{ip_{\phi}\phi}\xi(\beta_+),
\end{equation}
which represents a spreading wave packets in $\phi$ (times the wave function of the small anisotropy) due to the absence of any potential, since during the slow-rolling phase the Universe is on a potential plateau.
This can be considered as a starting point to understand that, by adding the dependence on space, its fluctuations can survive to the de Sitter phase producing the seeds for structure formation.

\section{Concluding remarks}\label{IX}
In this paper we analyzed the quasi-isotropization process of a Taub Universe in which the volume is quasi-classical and exponentially expands during a de Sitter phase, while the anisotropy degree of freedom is treated on a pure quantum level. We included into the dynamics also a massless and a minimally coupled scalar field, analyzed first as a classical field which contributes through its energy to the volume dynamics and then on a quantum level, like the anisotropy variable.
This field mimics here the contribution of the kinetic term of the inflaton field during a slow-rolling phase, when its potential energy is well-summarized by the cosmological constant term \cite{Primordial, Early}. 
	
The main merit of the present analysis consists of a detailed characterization of the quantum anisotropy decaying, as an effect of the exponential expansion of the Universe volume, here behaving as an external clock \cite{Vilenkin, Kiefer1991}.
Actually, we solved a time-dependent Schr\"odinger equation for the anisotropy quantum degree of freedom, analyzing the behaviour of Gaussian packets, both as expanded in the basic problem eigenfunctions, as well as exact states of the quantum dynamics. We see that the variance of the anisotropy variable decreases to a finite small value as the expansion goes by. Hence, a crucial point is that it survives after the de Sitter phase, differently from its classical behaviour described by pure classical Hamilton equations and in which the spatial curvature is negligible with respect to the cosmological constant term.
	
Furthermore, when we consider the scalar field as a quantum degree of freedom we see that its variance has a very different behaviour with respect to that one of the anisotropy. 
In fact, such a quantity spreads as the expansion goes by and 
this reflects the absence of a significant potential governing its dynamics during the slow-rolling phase. 
	
Here, we are considering a pure homogeneous field but its dynamics could be easily extended to the presence of inhomogeneous quantum corrections and it is clear that just the non-suppression of the scalar mode by the exponential expansion is the reason why it can generate seeds for later structure formation across the Universe.
	
On the contrary, we identified in the spatial curvature the ingredient responsible for the anisotropy quantum suppression. By other words, when the Universe can be characterized by small quantum anisotropies, in the sense discussed in \cite{Vilenkin} and in \cite{Mont-Agost},
the scalar potential takes the form of a harmonic oscillator which frequency increases with time as the Universe volume expands. This potential term is then responsible for the damping of the anisotropy.
	
The validity of this picture has to be regarded as viable on a rather general setting also in the presence of local inhomogeneities in the Universe. This offers an intriguing paradigm for the emergence of a classical and quasi- homogeneous ( a part from a relic qantum anisotropy) Universe from a primordial quantum age. 
	
In this respect, a crucial question calls now attention to be investigated: how the full quantum Universe can spontaneously evolves to the proposed picture a la Vilenkin, when its volume is a quasi-classical variable and the anisotropies are small. 
An answer to this highly non-trivial question probably requires to account for the presence of a Universe radiation component, able to alter the mixmaster dynamics in such a way that central regions of the Bianchi IX potential (though also in a local inhomogeneous scenario) are favoured, see \cite{Belinski2014} and reference therein.


\end{document}